\documentclass[draft,aps,prl,preprint,showpacs,amsmath,amssymb]{revtex4-1}

\begin{document}


\title{Spectral anti-broadening due to four-wave mixing in optical fibers}

\author{Alexander M. Balk}
\affiliation{Department of Mathematics, University of Utah, Salt Lake City, Utah, USA}
\email{balk@math.utah.edu} 



\date{\today}

\begin{abstract}
We show that the four-wave mixing can restrict spectral broadening.
This is a general physical phenomenon that occurs 
in \emph{one-dimensional} systems of four wave packets that resonantly interact ``2-to-2'':
$ \omega_1+\omega_2=\omega_3+\omega_4,\; k_1+k_2=k_3+k_4$, 
when an annihilation of one pair of waves results in the creation of another pair.
In addition, for this phenomenon to occur,
the group velocities $C_1,C_2,C_3,C_4$ of the packets should be in a certain order: 
The extreme value (\emph{max} or \emph{min}) of the four group velocities should be 
in the same pair with the middle value of the remaining three, e.g. $C_1<C_3<C_2<C_4$.
This phenomenon is due to the presence of an extra 
invariant, in addition to the energy, momentum, and Manley-Rowe relations.  
\end{abstract}


\maketitle 

\section{Introduction}
\label{Sect:Intro}
Usually nonlinear interaction leads to the spectral broadening:
When a wave packet propagates in an optical fiber, 
its energy spectrum $E_\omega$ ``spreads'' 
and becomes a broader distribution of energy over frequencies. 
Sometimes spectra broaden tens of times \cite{Agrawal13}.
Obviously, it depends on a particular situation, 
which nonlinearity is the main contributor to the spectral broadening. 
For instance, the paper \cite{Churkin}  argues 
that the four-wave mixing (FWM) is the main mechanism of spectral broadening 
in long Raman fiber lasers.
The goal of the present paper is to show that FWM, under certain conditions, 
can also act in the opposite way 
and actually restrict spectral broadening. 
This phenomenon is due to the presence of an extra invariant, 
in addition to the energy, momentum, and Manley-Rowe relations.

\section{Extra Invariant}
\label{Sect:Extra}
We consider a system of four resonantly interacting wave packets propagating in an optical fiber.
Each packet consists of waves with frequencies $\omega_j$ around $\omega^0_j$ 
and wave numbers $k_j=\beta^j(\omega_j)$ around $k^0_j$ ($j=1,2,3,4$). 
The basic frequencies $\omega^0_j$ and wave numbers $k^0_j$ are in resonance
\begin{equation}\label{Reso0}
\omega^0_1+\omega^0_2=\omega^0_3+\omega^0_4,\quad
k^0_1+k^0_2=k^0_3+k^0_4.
\end{equation}
The waves, that compose the packets, form many resonant quartets
\begin{eqnarray}\label{Reso}
\omega_1+\omega_2 &=& \omega_3+\omega_4,\nonumber\\
\beta^1(\omega_1)+\beta^2(\omega_2) &=& \beta^3(\omega_3)+\beta^4(\omega_4).
\end{eqnarray}
It does not matter for our purposes whether the waves belong to physically different modes 
--- with different functions $\beta^j(\omega)$ --- or to the same mode --- 
with all or some $\beta^j(\omega)$ being the same function; 
however, if some packets belong to the same branch, 
we require that they are well separated (do not overlap) in frequencies.

The resonance conditions (\ref{Reso}) express the energy and momentum conservation 
\begin{eqnarray}
   \int \omega N^1_\omega d\omega\;
+\;\int \omega N^2_\omega d\omega 
+\;\int \omega N^3_\omega d\omega+\int \omega N^4_\omega d\omega\; 
&=&\mbox{ const},\label{Energy}\\
   \int \beta^1(\omega) N^1_\omega d\omega\;
+\;\int \beta^2(\omega) N^2_\omega d\omega 
+\;\int \beta^3(\omega) N^3_\omega d\omega\;
+\;\int \beta^4(\omega) N^4_\omega d\omega\; 
&=&\mbox{ const},\label{Momentum}
\end{eqnarray}
where $N^j_\omega$ is the wave action spectrum for the $j$-th wave packet, 
or the number of photons of the $j$-mode with frequency $\omega$.

One can ask if there are other invariants
\begin{eqnarray}\label{Invar}
   \int \phi^1(\omega) N^1_\omega d\omega\;
+\;\int \phi^2(\omega) N^2_\omega d\omega\; 
+\;\int \phi^3(\omega) N^3_\omega d\omega\;
+\;\int \phi^4(\omega) N^4_\omega d\omega\; 
=\mbox{const},
\end{eqnarray}
with a different set of functions $\phi^j(\omega)$ 
--- other than the ones in (\ref{Energy}) and (\ref{Momentum}).
The invariant (\ref{Invar}) would take place 
if there were four functions $\phi^j(\omega)$ 
such that the resonance conditions (\ref{Reso}) implied another relation \cite{ZSch0}
\begin{eqnarray}\label{InvEq}
 \phi^1(\omega_1)+\phi^2(\omega_2)=\phi^3(\omega_3)+\phi^4(\omega_4).
\end{eqnarray}
The equation (\ref{InvEq}) is supposed to be linear independent 
of the resonance equations  (\ref{Reso}),
so that the conservation (\ref{Invar}) is essentially new --- independent 
of the energy-momentum conservation (\ref{Energy})-(\ref{Momentum}).
Each of the Manley-Rowe relations provides such conservation, e.g.
\begin{eqnarray*}
  \int N^1_\omega d\omega+
                         \int N^3_\omega d\omega =\mbox{const}\qquad\mbox{or}\qquad\\ 
\int N^1_\omega d\omega+\int N^2_\omega d\omega
+\int N^3_\omega d\omega+\int N^4_\omega d\omega =\mbox{const};
\end{eqnarray*}
they correspond to the obvious relations $1+0=1+0$ or $1+1=1+1$.

If the functions $\beta^j(\omega)$ are linear, 
then there is one more relation \cite{BaFe} of the kind (\ref{InvEq}).  
This situation is realized when the wave packets are narrow enough, so that 
we can approximate their dispersion relations by linear functions
\begin{eqnarray*}\label{p}
 k_j=\beta^j(\omega_j)=k^0_j+a_j \; p_j, \mbox{ where } p_j=\omega_j-\omega^0_j,
\end{eqnarray*}
with constant coefficients
\begin{eqnarray}\label{a}
a_j=\frac{\partial \beta^j}{\partial \omega}(\omega^0_j),
\end{eqnarray}
which are inverses of the group velocities $C_j$. 
Since the basic frequencies $\omega^0_j$ and wave numbers $k^0_j$ are in resonance (\ref{Reso0}),
the relations (\ref{Reso}) take the form
\begin{equation}\label{Res}
p_1+p_2=p_3+p_4,\quad
a_1\, p_1+a_2\, p_2=a_3\, p_3+a_4\, p_4.
\end{equation}
The equations (\ref{Res}) imply the relation
\begin{equation}\label{bInvEq}
b_1\, p_1^2\,+\,b_2\, p_2^2\,=\,b_3\, p_3^2\, +\,b_4\, p_4^2
\end{equation}
with coefficients
\begin{eqnarray}\label{b}
b_1&=&(a_2-a_1)(a_3-a_1)(a_4-a_1),\nonumber\\
b_2&=&(a_1-a_2)(a_3-a_2)(a_4-a_2),\nonumber\\
b_3&=&-(a_1-a_3)(a_2-a_3)(a_4-a_3),\nonumber\\
b_4&=&-(a_1-a_4)(a_2-a_4)(a_3-a_4).
\end{eqnarray}
It is easy to check (\ref{bInvEq}): 
Just solve the linear equations (\ref{Res}), say with respect to $p_3$ and $p_4$,
substitute this solution into (\ref{bInvEq}), and observe an identity, 
which holds for arbitrary $p_1$ and $p_2$.

The relation (\ref{bInvEq}) means the presence of an extra invariant
\begin{eqnarray}\label{bInvar}
  I\equiv b_1 \int p^2\, N^1_p\, dp\;
+\;b_2 \int p^2\, N^2_p\, dp\; 
+\;b_3 \int p^2\, N^3_p\, dp\;
+\;b_4 \int p^2\, N^4_p\, dp\; 
&=&\mbox{ const}
\end{eqnarray}
in addition to the energy, momentum, and Manley-Rowe relations. 

\section{Spectral anti-broadening}
\label{Sect:Anti}
We will assume special ordering of the four coefficients $a_1,a_2,a_3,a_4$: 
The extreme (minimal or maximal) coefficient should be in pair  
with the middle coefficient of the remaining three. 
(These two coefficients should correspond to the pair of wave packets on the same side of 
the resonance equations (\ref{Res})). 
Then all coefficients (\ref{b}) have the same sign, 
and the invariant $I$, defined in (\ref{bInvar}), is sign-definite. 
For example, $a_1<a_3<a_2<a_4$;
then all the coefficients (\ref{b}) are positive, 
and the extra invariant $I$ is positive-definite. 
Another example is $a_2>a_3>a_1>a_4$; 
then all the coefficients (\ref{b}) are negative, 
and the extra invariant $I$ is negative-definite.

The goal of the present paper is to show that 
there is restriction on spectral broadening of the wave packets.
The presence of the sign-definite invariant (\ref{bInvar}) shows this restriction.
 
Indeed, suppose on the contrary, that some wave packet, say \# 1, broadens;
some of its energy is transferred away from the basic frequency $\omega^0_1$ 
(i.e.\ towards bigger $|p_1|$); 
and the first integral in (\ref{bInvar}) increases. To keep the quantity $I$ constant, 
the other wave packets (\# 2, 3, and 4) can shrink (become less broad). 
However, they can make only finite amount of the quantity $I$ available 
for the increase of the first integral;
even if they shrink to zero, and the corresponding three integrals in (\ref{bInvar}) vanish, 
it is probably not enough to compensate for a significant increase of the first integral. 

This argument also includes restriction on shifting 
(or any deviation from the original basic frequency $\omega^0_1$).

Let us consider a simple model example, when the wave packets are Gaussian
\begin{equation*}
N^j_p=F_j \exp\left[-\frac{(p-\bar p_j)^2}{2\sigma_j^2}\right];
\end{equation*}
each packet is centered at $\bar p_j$, has width $\sigma_j$, and possesses total energy 
\begin{equation*}
E_j=\int [\omega^0_j+ p] N^j_p\, dp=(\omega^0_j+\bar p_j) F_j \sqrt{\pi} \sigma_j.
\end{equation*}
The invariant (\ref{bInvar}) is
\begin{eqnarray}\label{GaussI}
I&\equiv& I_1+I_2+I_3+I_4=\mbox{const}, \\ 
&&\mbox{where}\quad
I_j=b_j \int p^2 N^j_p\, dp=\frac{b_j(2 \bar p_j^2+\sigma_j^2)}{2(\omega^0_j+\bar p_j)} E_j.\nonumber
\end{eqnarray}
In the course of the FWM, the parameters $\bar p_j,\sigma_j,F_j$ change. 
Originally, all $\bar p_j=0$ (because of the choice of $\omega^0_j$).
Let us assume that all the ratios $b_j/\omega^0_j$ are roughly the same
and neglect $\bar p_j$ compared to $\omega^0_j$.
Since all the quantities $I_j$ are positive, 
the conservation (\ref{GaussI}) requires that 
\begin{eqnarray*}
(2 \bar p_1^2+\sigma_1^2)E_1 \lesssim
\sigma_{10}^2E_{10}+\sigma_{20}^2E_{20}+\sigma_{30}^2E_{30}+\sigma_{40}^2E_{40},
\end{eqnarray*}
where $\sigma_{j0}$ and $E_{j0}$ are the original width and energy of the four wave packets.
This inequality shows restriction on spectral broadening $\sigma_1$ and shifting $\bar p_1$ 
of the first wave packet, unless its energy $E_1$ becomes small. 

This counter-intuitive phenomenon of anti-broadening is actually similar 
to the energy transfer in hydrodynamic turbulence. 
In the 3-dimensional (3D) turbulence,
there is energy flux from large-scale eddies to small-scale eddies. 
However in the 2-dimensional (2D) turbulence, the energy flux is reversed: 
The energy flows from small-scale eddies towards large-scale eddies, 
sometimes accumulating in a single large-scale vortex. 
The well known reason for this phenomenon is 
the presence of an additional positive-definite invariant --- 
{\it enstrophy}; this invariant is present in 2D, but absent in 3D. 
The enstrophy restricts the energy transfer towards small scales. 

Similar behavior occurs in the interaction of four wave packets.
Usually, nonlinearity leads to the spectral broadening, 
i.e.\ to the energy flux towards small scales (large $|p_j|=|\omega_j-\omega^0_j|$).
When the system is 1-dimensional and the wave packets are narrow enough, 
there is an extra invariant (\ref{bInvar}).
With the special ordering, this invariant is sign-definite.
It restricts the energy transfer towards small scales, 
and the energy flux could reverse its direction, 
so that the energy would flow towards large scales (small $|p_j|$). 

Many different waves (with non-zero $p_j$) make resonant quartets and exchange energy.
The extra invariant provides book-keeping of 
the energy transfers back and forth between the four wave packets and shows
the final result of anti-broadening.

If a wave packet becomes less broad (more monochromatic), 
its modulational instability becomes more effective \cite{Soljacic00}. 
So, the modulational instability acts similar 
to the large-scale dissipation in the hydrodynamic turbulence.

Finally, let us note that the ordering of the coefficients (\ref{a}), 
stated in the beginning of the current Section,
is equivalent to the similar ordering of the group velocities $C_j=1/a_j$:
The extreme value of the four group velocities should be in pair 
with the middle value of the remaining three velocities. 
(The pair of values should correspond to the wave packets on the same side 
of the resonance equations (\ref{Reso}).)
Indeed,
\begin{eqnarray*}
 0<a_1<a_3<a_2<a_4\quad&\Leftrightarrow&\quad 0<C_4<C_2<C_3<C_1,\\
 a_1<0<a_3<a_2<a_4\quad&\Leftrightarrow&\quad C_1<0<C_4<C_2<C_3,\\
 a_1<a_3<a_2<0<a_4\quad&\Leftrightarrow&\quad C_2<C_3<C_1<0<C_4,\\
 a_1<a_3<a_2<a_4<0\quad&\Leftrightarrow&\quad C_4<C_2<C_3<C_1<0.
\end{eqnarray*}

\section{Conclusion}
We have seen that the FWM can impose restriction on the spectral broadening,
 contrary to our usual expectations. 
This is due to the presence of the extra invariant. The phenomenon 
occurs in the system of four wave packets (satisfying the phase matching condition), 
when the group velocities $C_1$ and $C_2$ of two annihilated waves 
and the group velocities $C_3$ and $C_4$ of two created waves are in certain order: 
The {\it extreme} value ({\it max} or {\it min}) of the four group velocities should be in pair with 
the {\it middle} value of the remaining three group velocities, e.g.\ $C_1<C_3<C_2<C_4$;
then the extra invariant is sign-definite.

Natural questions arise: Does the anti-broadening \\
(I) only slows down the spectral broadening or \\
(II) in some situations actually leads to spectral {\it narrowing}, 
similar to the inverse cascade in the hydrodynamic turbulence? 
(See the discussion of this analogy in the previous Section.)\\
Are there physically interesting situations where
the anti-broadening overcomes broadening by other nonlinearities?\\
Anyway, we see that if the FWM can be significant, 
one can spectrally broaden the wave packets more effectively 
if he avoids the above ordering.
On the other hand, sometimes the goal is to have spectrally narrow wave packets,
then the above ordering would be helpful.



%
%

%


\bibliography{My}

\end{document}